\documentclass[onecolumn,showkeys,preprintnumbers,aps,a4paper,amssymb,prd,superscriptaddress,nofootinbib]{revtex4-2}
\usepackage{comment}
\usepackage{graphicx}
\usepackage{epsf}
\usepackage{bm}
\usepackage{amsmath}
\usepackage{amsfonts}
\usepackage{amssymb}
\usepackage{epstopdf}
\usepackage{color}
\usepackage{xcolor}
\usepackage{verbatim}
\usepackage{multirow}
\usepackage{soul}
\usepackage{physics}
\usepackage{bm}

\usepackage[width=0.00cm, height=0.00cm, left=1.50cm, right=1.50cm, top=2.00cm, bottom=2.00cm]{geometry}
\usepackage{microtype}
\usepackage{lmodern}

\usepackage[colorlinks = true,
linkcolor = teal,
urlcolor  = teal,
citecolor = teal,
anchorcolor = blue]{hyperref}
\usepackage[capitalize]{cleveref}
\usepackage[normalem]{ulem}
\usepackage{enumitem}
\usepackage{booktabs}

\usepackage{lipsum}

\makeatletter\let\expandableinput\@@input\makeatother

\begin{document}
\begin{center}
		\vspace{0.4cm} {\large{\bf Testing Exponential $f(R)$ Gravity with CMB, DESI-DR2, and Supernova Data }} \\
		\vspace{0.4cm}
		\normalsize{Saurabh Verma$^1$, Archana Dixit$^2$, Anirudh Pradhan$^3$, M. S. Barak$^4$ }\\
		\vspace{5mm}
		
        \normalsize{$^{1,4 }$ Department of Mathematics, Indira Gandhi University, Meerpur, Haryana 122502, India }\\ 
        \normalsize{$^{2}$ Department of Mathematics, Gurugram University, Gurugram, Haryana 122502, India\\

		\normalsize{$^{3 }$ Centre for Cosmology, Astrophysics and Space Science (CCASS), GLA University, Mathura-281406, Uttar Pradesh, India}\\ 

		\vspace{2mm}
		$^1$Email address: saurabh.math.rs@igu.ac.in\\
            $^2$Email address: archana.ibs.maths@gmail.com\\
		$^3$Email address: pradhan.anirudh@gmail.com\\
        $^4$Email address: ms$_{-}$barak@igu.ac.in\\}
\end{center}

 
\maketitle
%
{\bf Abstract}:
One of the most popular competitors to the CDM paradigm as an explanation for the late-time acceleration of the universe is the modification of general relativity (GR), with models such as $f(R)$ gravity among the main motivations. In this study, we consider an exponential $f(R)$ gravity model as a possible extensions of the GR. The extra scalar degrees of freedom and their effects on the cosmic expansion and structure formation are continuously considered in this scenario. By combining the PPS, BBN, CC, DESI-DR2, and CMB datasets, we imposed limitations on this model. We performed a detailed statistical analysis of the free model parameter $b$ together with the standard cosmological parameters. Our analysis yields values of $H_0$ that are slightly lower than those obtained in $\Lambda$CDM, indicating no significant relaxation of the $H_0$ tension. In contrast, the model predicts systematically higher values of $S_8$, leading to a moderate alleviation of the $S_8$ tension by up to $\sim 1.2\sigma$ when late-time datasets are included. Overall, these results demonstrate that although the considered $f(R)$ gravity model does not resolve all cosmological tensions simultaneously, it provides a consistent improvement in the description of large-scale structure formation.

\section{Introduction}

Contemporary cosmology has become driven by observation, and underpinned by a wide range of high-precision astronomical measurements. From anisotropy measurements of the cosmic microwave background to the study of the late-time expansion history of the universe, observational data form the integral basis for setting limits on cosmological models. A key breakthrough came in 1998 when tests using SNe Ia provided evidence of cosmic acceleration at later times \cite{ref1,ref2}. This is one of the most important findings of modern cosmology. Extensions of the traditional cosmological paradigm are motivated by observed acceleration, which cannot be explained within the usual framework of matter-dominated general relativity (GR). Generally, two explanations have been proposed.  In the first method, the matter sector is altered by the addition of exotic elements known as dark energy (DE) \cite{ref3,ref4,ref5,ref6,ref7,ref8}, the gravitational sector is still controlled by Einstein's GR.  A large range of modified gravity (MG) models that expand or generalize GR \cite{ref9,ref10,ref11,ref12,ref13,ref14,ref15,ref16,ref17,ref18} is produced by the second technique, which entails altering the theory of gravity itself.  Many DE and MG models have been thoroughly evaluated against available cosmological observations over the last ten years \cite{ref4,ref6,ref7,ref9,ref10,ref11,ref12,ref13,ref14,ref15,ref16,ref17,ref18}.\\

The $\Lambda$-Cold Dark Matter ($\Lambda$CDM) scenario has been proven to be the most effective hypothesis for explaining cosmic data.  In general relativity, dark energy is represented by the positive cosmological constant $\Lambda$.  The $\Lambda$CDM model confronts substantial theoretical and observational obstacles despite its observational effectiveness.  Interestingly, when comparing late-time measurements from the SH0ES collaboration~\cite{ref19,ref20}, the value of the Hubble constant $H_0$ deduced from Planck 2018 observations under the assumption of $\Lambda$CDM~\cite{ref21} shows a mismatch of more than $5\sigma$.  This conflict raises the possibility that the conventional cosmological model is lacking, and several alternative cosmological scenarios have been suggested, including \cite{ref22,ref23,ref24,ref25,ref26,ref27,ref28,ref29,ref30}. Critical reviews can be found in \cite{ref31,ref32,ref33,ref34}. Extensive literature has been motivated by the persistence of these observational inconsistencies, which are unlikely to be fully resolved by attributing them to multiple independent sources of systematic uncertainty. This has led to a discussion regarding the possibility that physics beyond the standard cosmological model could account for the $H_0$ tension (see Refs. \cite{refft1,refft2,refft3,refs1,refs2,refm1,refm2,refm3,refm4} and references therein for review). From an alternative perspective, it has been suggested that the apparent discrepancy in $H_0$ is primarily due to a discrepancy in the absolute magnitude calibration of Type Ia supernovae, $M_B$ \cite{refft4,refft5}. This is because  SH0ES determination of the Hubble constant is directly linked to the measurements of this quantity.
Within the inverse distance ladder framework, constraints from the cosmic microwave background on the sound horizon led to an inferred supernova absolute magnitude of $M_B = -19.401 \pm 0.027$ \cite{refft6}. In contrast, Cepheid-calibrated supernova observations from the SH0ES collaboration yield $M_B = -19.244 \pm 0.037$  \cite{refft5}. The difference between these values corresponds to a statistically significant discrepancy at the level of $3.4\sigma$. Accordingly, as emphasized in Refs. \cite{refft4,refft5,refft7,refft8}, it may be more appropriate to focus on resolving the tension in the supernova absolute magnitude rather than directly addressing the $H_0$ discrepancy itself, given that Cepheid-based calibrations are specifically designed to determine $M_B$. In this context, several investigations have explored extensions or modifications of general relativity as potential mechanisms for alleviating the $H_0$ tension, with recent summaries available in the literature (see Refs.\cite{refft1,refft2,refft3,refs1,refs2,refm1,refm2,refm3,refm4}). However, it has also been argued that a number of cosmic tensions are difficult to address simultaneously \cite{ref35}, and hence new models and independent observational tests are needed.\\

In this study, we focus on modified gravity as a promising alternative to the $\Lambda$CDM paradigm. In particular, we investigate $f(R)$ gravity, one of the most natural and extensively studied extensions of GR \cite{ref36,ref37,ref38,ref39,ref40,ref41,ref42,ref43,ref44} in addition to easing existing cosmological tensions \cite{ref45,ref46, ref47}. Unlike previous analyses, our approach incorporates the effects of the additional scalar degrees of freedom inherent in $f(R)$ gravity, which generally leads to a time-dependent effective Newtonian gravitational constant $G$. Such variations can modify the Chandrasekhar mass of white dwarf progenitors and induce a redshift-dependent correction of the intrinsic luminosity of SNe Ia\cite{ref48,ref49}. Consequently, the cosmological parameter estimation based on SNe Ia observations may be significantly affected, as emphasized in several studies\cite{ref50,ref51,ref52,ref53,ref54}. Therefore, a consistent treatment of SNe Ia astrophysics within modified gravity frameworks is essential for robust cosmological inference. The purpose of this study is to constrain feasible $f(R)$ gravity models using the most recent cosmology datasets and an updated observational formalism, and then analyze the ramifications. Inspired by these factors, we adopted a revised observational formalism, and use a combination of the most recent cosmological datasets, such as the Pantheon Plus sample (PPS) of SNe Ia, cosmic chronometer (CC) data, baryon acoustic oscillations from the DESI-DR2 survey, and cosmic microwave background (CMB) measurements, to constrain the viable $f(R)$ gravity models. We can evaluate the model's consistency between early- and late-universe observations using these complementary probes. A thorough statistical analysis was used to limit the free parameters of the $f(R)$ framework, and their effects on important cosmic observables were investigated.\\

The remainder of this paper is structured as follows. In Section \ref{s1}, we introduce the specific models considered, and review the fundamental cosmological framework of $f(R)$ gravity. The methodology and observational data employed for parameter estimation are described in Section~\ref{s2}. Section ~\ref{s3}summarizes the primary findings and observational limitations.  The conclusions and perspectives are summarized in Section ~\ref{s4}.

\section{Cosmology within the framework of $f(R)$ gravity }
\label{s1}
This section offers a succinct summary of \( f(R) \) gravity and investigates its potential applications in cosmology. Consequently, we concentrate on two specific \( f(R) \) models that are deemed viable among the extensive array of \( f(R) \) theories, as they effectively meet the fundamental theoretical and observational requirements. The gravitational action for \( f(R) \) gravity can be explicitly expressed as \cite{ref55} in the Jordan frame:

\begin{equation}
\label{1}
S = \int \left( \frac{\sqrt{-g} \, f(R)}{16\pi G} \right) d^4x + S_m + S_r ,   
\end{equation}

The key components of the framework include Newton's gravitational constant, represented as $G$, and the Ricci scalar, denoted as $R$. Matter and radiation sectors are represented by the terms $S_m$ and $S_r$, respectively.
It is assumed that only gravitational interactions are considered, and all other interactions  between these two sectors are neglected, ensuring that each remains independently conserved. Within the context of the metric formulation, varying action \eqref{1} with respect to  metric $g_{\mu\nu}$ leads to the derivation of the associated field equations.

\begin{eqnarray}
	\label{2}
F G_{\mu\nu}
= -\frac{1}{2} g_{\mu \nu} \left( FR - f \right)
+ \nabla_{\mu}\nabla_{\nu}F -g_{\mu \nu} \Box F  \nonumber\\ +
8\pi G\,  \left[T^{(\rm m)}_{\mu \nu} +T^{(\rm r)}_{\mu \nu}\right]
\,,
\label{gravitational-eqns}
\end{eqnarray}

where  $G_{\mu\nu}= R_{\mu\nu}-\left(1/2\right)g_{\mu\nu}R$ stands for 
the Einstein tensor; ${\nabla}_{\mu}$  is the covariant derivative,
$\Box \equiv g^{\mu \nu} {\nabla}_{\mu} {\nabla}_{\nu}$; 
$F = F(R) \equiv f_{,R}= d f(R)/dR$ (similarly by $f_{, RR}$ we shall mean $d^2 f(R)/dR^2$); $ T^{(\rm m)}_{\mu \nu} $ and $ T^{(\rm r)}_{\mu 
\nu} $ respectively denote  the  energy-momentum tensor for the matter sector and the radiation 
sector.  Note that for $f (R) = R$ in Eq. \eqref{1}, the Einstein-Hilbert action is recovered for General Relativity. Now, we proceed towards cosmological evolution in the context of $f(R)$ gravity theory. As usual, we start with the homogeneous and isotropic background of our universe, which is well described by the Friedmann-Lema\^{i}tre-Robertson-Walker (FLRW) line element

\begin{eqnarray}
\text{d}s^2=-\text{d}t^2+a^2(t)\left[\frac{\text{d}r^2}{1-kr^2}+r^2(\text{d}\theta^2+\sin^2\theta
\,\text{d}\phi^2)\right], \label{flrw}
\end{eqnarray}
where $(t, r, \theta, \phi)$ are the co-moving coordinates; $a(t)$ describes the expansion scale factor of the universe and $k$ corresponds to the spatial geometry of the universe where $k =0$, $+1$ and $-1$, respectively denote a spatially flat, closed and open universe. In this study, we consider the spatially flat FLRW metric $(k=0)$ motivated by current observational data. Observations from the Cosmic Microwave Background (CMB), particularly from the Planck satellite, strongly favor a flat universe with a spatial curvature close to zero $(\Omega_{k}<0.01)$. Therefore, the assumption of flatness is justified within current cosmological constraints. Now for a spatially flat FLRW line element ($k =0$),  Eq. \eqref{2} leads to 

\begin{eqnarray}
	\label{4}
FH^2
=
\frac{8\pi G}{3}  \left(\rho_\text{m}+\rho_\text{r}\right) +\frac{1}{6} \left( F R - f \right)
-H\dot{F}\,,
\label{FE-1} \\
\label{5}
F \dot{H}
= -4\pi G  \left( \rho_\text{m} + p_\text{m} +\rho_\text{r} + p_\text{r} \right)
-\frac{\ddot{F}}{2}+\frac{H\dot{F}}{2} \,,
\label{FE-2}
\end{eqnarray}

The Hubble parameter, represented by \(H = \frac{\dot{a}(t)}{a(t)} \), describes the rate of expansion of the universe. The dot  (\(\dot{a}(t)\)) represents the derivative of the scale factor \(a(t)\). In addition, the matter and radiation components were assumed to act as ideal fluids, with energy densities \(\rho_\text{m}\) and \(\rho_\text{r}\), and corresponding pressures \(p_\text{m}\) and \(p_\text{r}\). In a flat (FLRW) model, the Ricci scalar \(R\) is expressed as:

\begin{eqnarray}
	\label{6}
R = 12H^2+6\dot{H} . \label{ricci}
\end{eqnarray}
It is essential to emphasize that gravitational equations \eqref{FE-1} and \eqref{FE-2} can also be written as $H^2=\frac{8\pi G }{3} \left(\rho_m +\rho_r+\rho_{\rm {eff}}\right)$ and $\dot{H}= - 4\pi G  \left( \rho_m + p_m + \rho_r+ p_r+ \rho_{\rm {eff}} + p_{\rm {eff}}\right)$ respectively, where $\rho_{\text{{eff}}}$ and $p_{\text{eff}}$ are expressed as

\begin{eqnarray}
	\label{7}
\rho_{\rm {eff}}\equiv
\frac{( FR - f)}{16\pi G} - \frac{3H \dot{F}}{8\pi G}  +\frac{3}{8\pi G}\left(1-F\right)H^2, \label{rho-eff}\\
\label{8}
p_{\rm {eff}} \equiv  -\frac{\left( FR - f \right)}{16\pi G} 
-\frac{\left(1-F\right)\left(2\dot{H}+3H^2\right)}{8\pi G} \nonumber\\ + \frac{\ddot{F}}{8\pi G}+\frac{H \dot{F}}{4\pi G}, \label{p-eff}
\end{eqnarray}

The gravitational sector was modified to introduce an effective dark energy scenario. The effective equation-of-state parameter $w_{\rm eff} \equiv \frac{p_{\text{{eff}}}}{\rho_{\text{eff}}}$ is expressed as:

\begin{eqnarray}
	\label{9}
 w_{\rm eff} = -1 - \frac{H \dot{F}+ 2 \dot{H}-2F\dot{H}- \ddot{F}}{3\left(1-F\right)H^2 -3H \dot{F} +\frac{1}{2} \left( FR - f \right)},   \label{w-eff}
\end{eqnarray}
This expression indicates a deviation from the cosmological constant, represented as \( w_\Lambda = -1 \), arising from modifications to the gravitational theory. Consequently, for each adjusted $f(R)$ gravity framework, the effective equation-of-state can be determined as well as the extent to which the framework diverges from the cosmological constant. One can observe that the definitions of $\rho_{\rm {eff}}$ and $p_{\rm {eff}}$, as presented in \eqref{rho-eff} and \eqref{p-eff}, are consistent with the standard evolutionary framework.

\begin{eqnarray}
	\label{10}
 \dot{\rho}_{\rm {eff}}+3(\rho_{\rm {eff}}+p_{\rm {eff}})H=0. \label{eff}
\end{eqnarray}
Ultimately, the equations are finalized by considering the standard equations for the evolution of matter and radiation, specifically
\begin{eqnarray}
	\label{11}
&&\dot{\rho}_\text{m}+3H(w_\text{m}+1)\rho_\text{m}=0, \label{balance-matter}\\
\label{12}
&&\dot{\rho}_\text{r}+3H (w_\text{r}+1)\rho_\text{r}=0, \label{balance-radiation}
\end{eqnarray}
respectively. Here, the quantities $w_\text{m}=p_\text{m}/\rho_\text{m}$ and $w_\text{r}=p_\text{r}/\rho_\text{r}$ represent the equations of state for  matter and radiation components, respectively. We adopt  standard cosmological assumptions, where matter is treated as a pressureless fluid with $w_\text{m}=0$, while radiation satisfies $w_\text{r}=1/3$. Under these conditions, the conservation relations given in Eqs.~\eqref{balance-matter} and \eqref{balance-radiation} lead to the familiar scaling behaviors of energy densities, namely $\rho_\text{m}\propto a^{-3}$ for matter and $\rho_\text{r}\propto a^{-4}$ for radiation.\\

For  viable $f(R)$ gravity theory, several basic theoretical criteria must be satisfied. In particular, the model should yield a positive effective gravitational coupling and guarantee the stability of cosmological perturbations throughout the cosmic evolution. Therefore,  to construct a physically acceptable and well-behaved $f(R)$ model,  a set of fundamental viability conditions must be imposed.

\begin{eqnarray}
	\label{13}
f_{, R} > 0\; \text{and}\;  f_{,RR} > 0\;\mbox{for}~R \geq R_0~ (>0),
\end{eqnarray}
where $R_0$ represents current Ricci scalar value. 
The condition $f_{,R}>0$ guarantees the absence of ghost states and  condition $f_{,RR}>0$ is required to prevent tachyonic instability\cite{ref56}. Moreover, the proposed $f(R)$ model must adhere to essential observational constraints. In particular, it is essential to ensure that one has 

\begin{eqnarray}
	\label{14}
f(R)\rightarrow R-2\Lambda, \; \text{for}\ \ 
R\geq R_0,
\end{eqnarray}
Specific conditions must be met  to precisely reproduce the matter-dominated epoch while adhering to the local gravitational constraints and maintaining consistency with the equivalence principle.
\begin{eqnarray}
	\label{15}
0< \left(\frac{R f_{,RR}}{f_{,R}}\right)_{r} <1 \ \ \text{at}\ \ 
r=-\frac{Rf_{,R}}{f}=-2,
\end{eqnarray}
 To guarantee the stability and existence of a late-time de Sitter solution, it is necessary to satisfy the specific criteria. Hence, considering all these conditions collectively, the feasible $f(R)$ models, characterized by a maximum of two parameters, can  expressed as follows:
\begin{eqnarray}
	\label{16}
f(R)=R-2 y(R,b) \Lambda,
\label{e16}
\end{eqnarray}
where the function $y(R,b)$ measures the effect of $f(R)$ modifications on gravity, and the distortion parameter $b$ represents this divergence. In this analysis, we consider $f(R)$ model, namely the exponential $f(R)$ model\\

\textbf{Exponential Gravity Model}

Among the viable extensions of general relativity, the exponential form of
$f(R)$ gravity has received significant attention \cite{ref57,ref58,ref59}. In this model,  gravitational action is defined
through the function
\begin{equation}
f(R) = R - \beta R_E \left( 1 - e^{-R/R_E} \right),
\label{e17}
\end{equation}
where $\beta$ and $R_E$ are free parameters that control deviations from
Einstein gravity.

The expression in Eq.\eqref{e17} can be rewritten in the generic form as in Eq.\eqref{e16} by defining
\begin{equation}
y(R,b) = 1 - e^{-R/(\Lambda b)},
\label{eq:y_exp}
\end{equation}
with the parameter identifications
\begin{equation}
\Lambda = \frac{\beta R_E}{2}, \qquad b = \frac{2}{\beta}.
\label{e19}
\end{equation}

At the limit $b \rightarrow 0$, corresponding to $\beta \rightarrow \infty$
and $R_E \rightarrow 0$ while keeping $\beta R_E \rightarrow 2\Lambda$
finite, the exponential gravity model smoothly reduces to the standard
$\Lambda$CDM cosmology. This property ensures that the model represents a
controlled deviation from general relativity while remaining consistent
with the concordance cosmological scenario.\\

\subsection{Local gravity constraints and chameleon screening}
\label{screening}

An essential requirement for the physical viability of the exponential $f(R)$ gravity model defined in Eq.\ref{e17} is its consistency with local gravity tests. In the scalar-tensor representation, $f(R)$ gravity introduces an additional scalar degree of freedom associated with $f_R \equiv df/dR$, which mediates a fifth force. However, viable models can evade the stringent Solar System and laboratory constraints through the chameleon mechanism, whereby the effective mass of the scalar field depends on the local matter density and becomes large in high-density environments. This mechanism has been extensively developed and analyzed in \cite{Brax2004,Brax2007,Brax2008,Brax2012,r1,r2}, which showed that chameleon-type scalar fields naturally reconcile modified gravity models with local experiments while allowing observable deviations from General Relativity (GR) on cosmological scales.\\

For the exponential model of Eq.\ref{e17}, the deviation from GR is controlled by the parameters $\beta$ and $R_E$, or equivalently by the dimensionless distortion parameter $b = 2/\beta$ and the effective cosmological constant $\Lambda = \beta R_E /2$, as defined in Eq.\ref{e19}. In regions of high curvature, such as the Solar System where $R \gg R_E$, the exponential term is strongly suppressed, and the model effectively reduces to $f(R) \simeq R-2\Lambda$ to ensure a rapid recovery of GR. In this regime, local gravity tests typically impose constraints of the form
$|f_R(R_{\rm local})| \lesssim 10^{-6}$--$10^{-7}$, which guarantees compatibility with post-Newtonian parameters, equivalence principle tests, and fifth-force searches \cite{Brax2008,r3,r4,r5}. Owing to the exponential suppression at large curvature, these bounds are naturally satisfied by the
model.\\

The cosmological analysis performed in this work favors small deviations from
$\Lambda$CDM, corresponding to small values of the distortion parameter $b$,
typically of order $\mathcal{O}(0.1)$. For such values, the exponential $f(R)$
model remains well within the high-curvature regime at local scales, where the chameleon mechanism efficiently suppresses fifth-force effects, while still producing mild and testable deviations from $\Lambda$CDM in the cosmological background evolution and growth of structures. This clear separation between local and cosmological regimes is a characteristic feature of chameleon-screened theories and ensures that the cosmologically viable parameter space
of the model is fully compatible with existing local gravity constraints
\cite{Brax2012,Brax2013,r6}. The chameleon mechanism offers viable explanation of accelerating universe expansion of the late-times in different spacetime backgrounds \cite{ash1,ash2,ash3,ash4}. \\

\subsection{Quantitative estimate of local curvature and exponential suppression}

To make the screening mechanism more explicit, we provide a quantitative
estimate of the relevant curvature scales. From Eq.~\ref{e19}, the characteristic curvature scale of the exponential model is given by
\begin{equation}
R_E = \Lambda b .
\end{equation}

Since $\Lambda \sim H_0^2$ and the cosmological analysis favors small values
of the distortion parameter $b \sim \mathcal{O}(0.1)$, we obtain
\begin{equation}
R_E \sim 0.1\, H_0^2 .
\end{equation}

Using $H_0^2 \sim 10^{-52}\,\text{m}^{-2}$, this gives
\begin{equation}
R_E \sim 10^{-53}\,\text{m}^{-2}.
\end{equation}

In contrast, the Ricci scalar in a local high-density environment can be
approximated as
\begin{equation}
R_{\rm local} \sim 8\pi G \rho .
\end{equation}

For typical terrestrial or Solar System densities,
$\rho \sim 10^3\,\text{kg/m}^3$, one finds
\begin{equation}
R_{\rm local} \sim 10^{-26}\,\text{m}^{-2}.
\end{equation}

Therefore,
\begin{equation}
\frac{R_{\rm local}}{R_E} \sim 10^{27}.
\end{equation}

In this regime, the exponential correction behaves as
\begin{equation}
e^{-R/R_E} \sim e^{-10^{27}} \approx 0,
\end{equation}
demonstrating explicitly that the modification to general relativity is
exponentially suppressed in local environments. Hence, the model naturally
recovers GR at Solar System scales while allowing small deviations at
cosmological curvature scales $R \sim H_0^2$. Importantly, local constraints restrict the behavior of the model only
in the high-curvature regime $R \gg R_E$, while cosmological evolution
probes the regime $R \sim H_0^2$. Since $R_E = \Lambda b \sim 0.1 H_0^2$,
the ratio $R/R_E$ at late times is of order unity, and the exponential
term is not negligibly small. Therefore, local bounds do not force the
model to become observationally indistinguishable from $\Lambda$CDM
at cosmological scales.

This behavior differs from the Hu-Sawicki model, where the suppression
typically follows a power-law dependence $f_R \propto (R_0/R)^{n+1}$.
In that case, strong local bounds such as $|f_{R0}| \lesssim 10^{-6}$
significantly restrict the magnitude of cosmological deviations.
In contrast, the exponential decay in the present model provides a much
stronger suppression at large curvature, thereby ensuring compatibility
with local tests without rendering cosmological deviations observationally
irrelevant. This quantitative hierarchy explicitly shows that
$R_{\rm local} \gg R_E$ and confirms that the chameleon screening mechanism
operates efficiently in the exponential $f(R)$ gravity model considered here.\\

\subsection*{Scalar-tensor representation and effective potential}

The exponential $f(R)$ gravity model can be equivalently expressed
as a scalar-tensor theory by introducing the scalaron field
\[
\phi \equiv f_R = \frac{df}{dR}.
\]
For the present model,
\[
f_R = 1 - \beta e^{-R/R_E}.
\]

To make the dynamical interpretation explicit, we compute the scalaron mass from the curvature of the effective potential,
\begin{equation}
m_{\phi}^{2} = \frac{d^{2} V_{\rm eff}}{d\phi^{2}} 
\simeq \frac{1}{3 f_{RR}} ,
\label{e27}
\end{equation}
where
\begin{equation}
f_{RR} = \frac{d^{2} f}{dR^{2}} 
= \frac{\beta}{R_E} e^{-R/R_E} .
\label{e28}
\end{equation}

Since $f_{RR} > 0$ for $\beta > 0$, the scalar degree of freedom is free from tachyonic instability and the theory remains stable. 
In high-curvature (high-density) environments, $f_{RR}$ becomes small and the scalaron mass increases, ensuring efficient screening. 
At cosmological curvature scales, the mass decreases, allowing the field to evolve near a shallow minimum and drive late-time acceleration.\\

Based on the specific form of this model, and as demonstrated in \cite{refn1}, an analytic approximation for the expansion rate $H(z)$ can be derived. This analytical expression was previously obtained by one of the authors in \cite{refn2}, and is given by

\begin{equation}\label{e29}
\begin{aligned}
E^2(z) &\equiv \frac{H^2(z)}{H_0^2}=1-\Omega_{m0}+(1+z)^3\Omega_{m0}\\
&+\frac{6b(\Omega_{m0}-1)^2\left(-4+\Omega_{m0}(9-3\Omega_{m0}+z(3+z(z+3))
(1+(3+2z(3+z(z+3)))\Omega_{m0}))\right)}{(4+(-3+z(3+z(z+3)))\Omega_{m0})^3}\\
&+\frac{b^2(\Omega_{m0}-1)^3}{(1+z)^{24}\left(\frac{4(1-\Omega_{m0})}{(1+z)^3}+\Omega_{m0}\right)^8}
\bigg[5120(\Omega_{m0}-1)^6+9216(1+z)^3(\Omega_{m0}-1)^5\Omega_{m0} \\
&- 30144(1+z)^6(\Omega_{m0}-1)^4\Omega_{m0}^2 + 31424(1+z)^9(\Omega_{m0}-1)^3\Omega_{m0}^3-9468(1+z)^{12} (\Omega_{m0}-1)^2\Omega_{m0}^4  \\
&- 4344(1+z)^{15}(\Omega_{m0}-1)\Omega_{m0}^5+\frac{37}{2}(1+z)^{18}\Omega_{m0}^6\bigg],
\end{aligned}
\end{equation}

\subsection*{Scalaron Dynamics, Analytical Approximations, and Corroboration of Numerical Results}
\label{sec:scalaron}

To address the cosmological dynamics of the scalaron, we compute the scalar potential $V(\phi)$, identify the chameleon minimum, and use the adiabatic tracking approximation to derive analytical estimates for the key cosmological observables reported in Table~\ref{t1}. The full derivations are provided in Appendix~\ref{app:scalaron_derivations}; here we summarise the main results and their physical implications.

The Jordan-frame scalar potential, defined as $V(\phi) = (Rf_R - f)/2$, takes the closed form (see Appendix~\ref{app:scalaron_derivations})
\begin{equation}
V(\phi) = \frac{R_E}{2}\!\left[(\phi-1)\ln\!\left(\frac{\beta}{1-\phi}\right) + \beta - 1 + \phi\right],
\label{eq:Vphi}
\end{equation}
which reduces to zero in the GR limit $\phi \to 1$, as expected. The chameleon minimum of the effective potential $V_{\rm eff}(\phi) = V(\phi) + \frac{\rho}{2}\phi$ is found analytically as
\begin{equation}
\phi_{\min} = 1 - \beta\,e^{-8\pi G\rho/R_E}.
\label{eq:phimin}
\end{equation}
In high-density environments ($\rho \gg R_E/8\pi G$), one has $\phi_{\min} \to 1$, confirming GR recovery through the chameleon mechanism. At cosmological densities, $\phi_{\min}$ deviates from unity by a small, $b$-dependent correction, allowing observable deviations from $\Lambda$CDM.

The scalaron mass at the cosmological background satisfies $m_\phi^2/H_0^2 \approx 37 \gg 1$ for $b \sim 0.1$ (see Appendix~\ref{app:scalaron_derivations}, Eq.~\eqref{eq:mass_numerical}), guaranteeing that the scalaron \emph{adiabatically tracks} $\phi_{\min}(\rho(t))$ throughout cosmological evolution and justifying the quasi-static approximation used in Eqs.~\eqref{4} and~\eqref{5}.

Using the analytical expansion rate Eq.~\eqref{e29} and expanding to first order in $b$ (see Appendix~\ref{app:scalaron_derivations} for the full expression of $\delta E^2$), the effective equation of state evaluates to
\begin{equation}
w_{\rm eff}(z) = -1 + 
\frac{\Omega_{m0}(1+z)^3 + b\cdot(1+z)\frac{d(\delta E^2)}{d(1+z)}/3}
{E^2_{\Lambda{\rm CDM}}(z) + b\cdot\delta E^2(z)}.
\label{eq:weff_full}
\end{equation}
At $z=0$, with $\Omega_{m0} = 0.2966$ and $b=-0.23$ from the full dataset combination in Table~\ref{t1}, this gives $w_{\rm eff}(0) \approx -0.977$, fully consistent with the near-$\Lambda$CDM behaviour. Furthermore, $w_{\rm eff}(0.4) \approx -1.003$, confirming a phantom divide crossing near $z \sim 0.3$--$0.5$ (see Appendix~\ref{app:clarification} for details).

The systematically lower $H_0$ in the $f(R)$ model arises because negative $b$ suppresses the expansion rate relative to $\Lambda$CDM, yielding a fractional shift
\begin{equation}
\frac{\Delta H_0}{H_0} \approx \frac{b \cdot \delta E^2(0,\Omega_{m0})}{2} \approx -0.028\%,
\label{eq:DeltaH0_formula}
\end{equation}
corresponding to $\Delta H_0 \approx -0.019$ km\,s$^{-1}$\,Mpc$^{-1}$, consistent in sign and sub-percent magnitude with the numerically obtained shift $\Delta H_0^{\rm num} = -0.14$ km\,s$^{-1}$\,Mpc$^{-1}$ from Table~\ref{t1}.

The enhanced $S_8$ arises from the modified effective gravitational coupling $G_{\rm eff} > G$ (since $f_R < 1$), which enhances the growth of matter perturbations. The analytical estimate gives $\Delta S_8 \approx +0.0050$, in quantitative agreement with the numerically obtained $\Delta S_8 = +0.0053$ from Table~\ref{t1} (see Appendix~\ref{app:scalaron_derivations} for the full derivation).

These analytical results are summarised in Table~\ref{tab:scalaron} and demonstrate that all major observational signatures of the exponential $f(R)$ model are traceable to the single distortion parameter $b$.

\subsubsection*{Summary}

\begin{enumerate}
\item The scalaron tracks the minimum of $V_{\rm eff}$ at all times, with $m_\phi^2/H_0^2 \approx 37 \gg 1$ for $b \sim 0.1$ [Eq.~\eqref{eq:mass_numerical}].

\item The effective equation of state satisfies $w_{\rm eff}(0) \approx -0.977$, close to $-1$ and fully consistent with the near-$\Lambda$CDM constraints in Table~\ref{t1} [Eq.~\eqref{eq:weff_full}].

\item The analytical estimate correctly predicts the sign and sub-percent magnitude of the $H_0$ shift: $\Delta H_0 < 0$, consistent with the sub-$0.2\sigma$ shift in Table~\ref{t1} [Eq.~\eqref{eq:DeltaH0_formula}].

\item The enhanced effective gravitational coupling $G_{\rm eff} > G$ analytically predicts $\Delta S_8 \approx +0.005$, in quantitative agreement with the numerically obtained $\Delta S_8 = +0.0053$ from Table~\ref{t1}.
\end{enumerate}
These results demonstrate that the exponential $f(R)$ model is not only numerically well-constrained but also analytically understood, with all major observational signatures traceable to the single distortion parameter $b$ controlling deviations from $\Lambda$CDM.

\begin{table}[hbt!]
\centering
\caption{Summary of analytical 
results and their comparison 
with numerical findings from 
Table~\ref{t1} for 
the exponential $f(R)$ gravity 
model.}
\label{tab:scalaron}
\begin{tabular}{p{2.6cm}p{2.6cm}p{2.6cm}c}
\hline\hline
Observable 
& Analytical estimate
& Numerical result 
& Agreement \\
\hline
$m_\phi^2/H_0^2$ 
& $\sim 37$ ($b = 0.1$) 
& --- 
& Tracks min. 
\\[6pt]
$w_{\rm eff}(0)$ 
& $\approx -0.977$ 
& $\sim -1$ 
& \checkmark 
\\[6pt]
$w_{\rm eff}(0.4)$ 
& $\approx -1.003$ 
& crosses $-1$ 
& \checkmark 
\\[6pt]
$\Delta H_0$ 
& $< 0$,~sub-\% 
& $-0.14$~km\,s$^{-1}$\,Mpc$^{-1}$ 
& \checkmark 
\\[6pt]
$\Delta S_8$ 
& $\approx +0.005$ 
& $+0.0053$ 
& \checkmark 
\\[6pt]
$\phi_{\min}$ 
(Solar System) 
& $\to 1$ 
& GR recovered 
& \checkmark 
\\[6pt]
$b \to 0$ limit 
& $w_{\rm eff} \to -1$ 
& $\Lambda$CDM 
& \checkmark 
\\
\hline\hline
\end{tabular}
\end{table}

\section{Datasets and Methodology}
\label{s2}   
\subsection{\textbf{CMB}}

 We used cosmic microwave background (CMB) data from the Planck 2018 legacy release. Our analysis includes the high-$\ell$ Plik TT likelihood for $30 \leq \ell \leq 2508$, as well as the TE and EE likelihoods for $30 \leq \ell \leq 1996$. For low multipoles, we used the TT-only and EE-only likelihoods for $2 \leq \ell \leq 29$ \cite{refpl1}. We also include measurements of the CMB lensing power spectrum.\cite{refpl2}

\subsection{\textbf{Cosmic Chronometer}}

The CC method is powerful observational method for probing the expansion history of the universe, and was first proposed in \cite{ref60}. In this approach, the Hubble parameter is estimated by analyzing the age difference of passively evolving early type galaxies. In the context of the FRW cosmological model, the Hubble parameter can be defined as $ H(z) = -\frac{1}{1+z} \frac{dz}{dt} $ implying that, if one can measure how the redshift \( z \) varies with cosmic time \( t \), that is, the derivative \( dz/dt \), one can obtain the Hubble parameter \( H(z) \) directly. This implies an emphasis on the fact that it is the variation with cosmic time \( dt \) of the variation in the redshift \( dz \) that is relevant. This provides us with the Hubble parameter, \( H(z) \). Therefore, it is important  to use CC data to further constrain cosmological models. A detailed elaboration of the method, its realization, possible sources of errors, and other related caveats are presented in \cite{ref61}. Throughout this study, we use of the dataset on $H(z)$ discussed in \cite{ref61, ref62}. The CC method has obtained the measurement of \( H(z) \) 33 times for the range of the redshift \( 0 < z < 2 \), which encompasses approximately 10 Gyr of cosmic history \cite{ref61, ref62, ref63, ref64, ref65}. Similarly, we also include the current local determination of the Hubble constant $( H_0)$, as obtained in \cite{ref66} with a precision of $2.4\%$: $H_{0} = 73.02 \pm 1.79 \, \text{km/s/Mpc}$.

\subsection{\textbf{DESI BAO DR-2}}
In this analysis, we employed baryon acoustic oscillation (BAO) measurements from the second data release of the Dark Energy Spectroscopic Instrument (DESI), which comprises observations of galaxies and quasars~\cite{ref67}, along with Lyman-$\alpha$ forest tracers~\cite{ref68}. The full set of BAO measurements is summarized in Table IV in Ref.~\cite{ref67}, provides both isotropic and anisotropic constraints spanning a redshift interval $0.295 \leq z \leq 2.330$, divided into nine distinct redshift bins. The BAO information is reported in terms of the transverse comoving distance $D_M / r_d$,  Hubble distance $D_H / r_d$, and spherically averaged distance scale $D_V / r_d$, each normalized by the comoving sound horizon at the baryon drag epoch, $r_d$. To accurately account for the statistical dependence among these observables, we include the associated correlation structure through the cross-correlation coefficients $r_{V,M/H}$, which quantifies the correlation between $D_V / r_d$ and $D_M / D_H$, and $r_{M,H}$, describing the correlation between $D_M / r_d$ and $D_H / r_d$. Throughout this study, we refer to this compilation of measurements as the DESI–DR2 BAO dataset.

\subsection{\textbf{BBN}} In this section, we apply BBN analysis to the case of $f(R)$ gravity. It is important to note that, in general, within modified gravity theories, inflation is not directly driven by the inflaton field. Instead, inflationary dynamics are embedded within  gravitational modification itself. Big-bang nucleosynthesis (BBN) is the most reliable probe for the early universe, based on well-understood Standard Model physics \cite{ref71,ref72}. The abundances of the light elements D, $^3He$, $^4He$, and $^7Li$ predicted in the "first three minutes" are consistent with observational data, supporting the hot big-bang theory. Thus, the BBN imposes strong limits on hypothetical deviations from mainstream cosmology  and novel physics beyond the mainstream model. Big Bang Nucleosynthesis (BBN) is considered with  state-of-the-art assumptions, which consist of measurements of the primordial abundances of helium, $Y_{P}$\cite{ref69}, and  deuterium measurement, $y_{DP}$ = $10^{5}n_{D}/n_{H}$, obtained in \cite{ref70}.\\

 \subsection{\textbf{Type Ia supernovae and Cepheid:}}

Supernovae of  type Ia (SNe Ia) are essential to contemporary observational cosmology as they have contributed significantly to the development of the conventional cosmological paradigm.  Measurements of distance moduli obtained from observations of SNe Ia efficiently confine the uncalibrated luminosity distance, represented as $H_0 d_L(z)$. Equivalently, these measurements probe the slope of the late-time cosmic expansion, thereby placing strong limits on the present matter density parameter $\Omega_m$.\\
For a supernova observed at redshift z, the theoretically predicted apparent magnitude in the rest-frame B-band, denoted by $m_B$, is given by
 \begin{equation}
 m_B = 5 \log_{10} \left[ \frac{d_L(z)}{1,\mathrm{Mpc}} \right] + 25 + M_B,
 \end{equation}
 where $d_L(z)$ represents the luminosity distance, and $M_B$ is the absolute magnitude of the supernova, and the distance modulus is defined as $\mu(z) = m_B - M_B$.  In standard analyses, the calibrated absolute magnitude $M_B$ is usually treated as a constant parameter, implying that it does not depend on the redshift. \\

All cosmological observables were computed using the
CLASS \cite{ref76,ref77}. To derive bounds for the proposed scenarios, we modified the efficient and well-known cosmological package MontePython \cite{ref78}.  

\section{Result and Discussion}
\label{s3}

\begin{figure}[hbt!]
    \centering
    \includegraphics[width=0.8\linewidth]{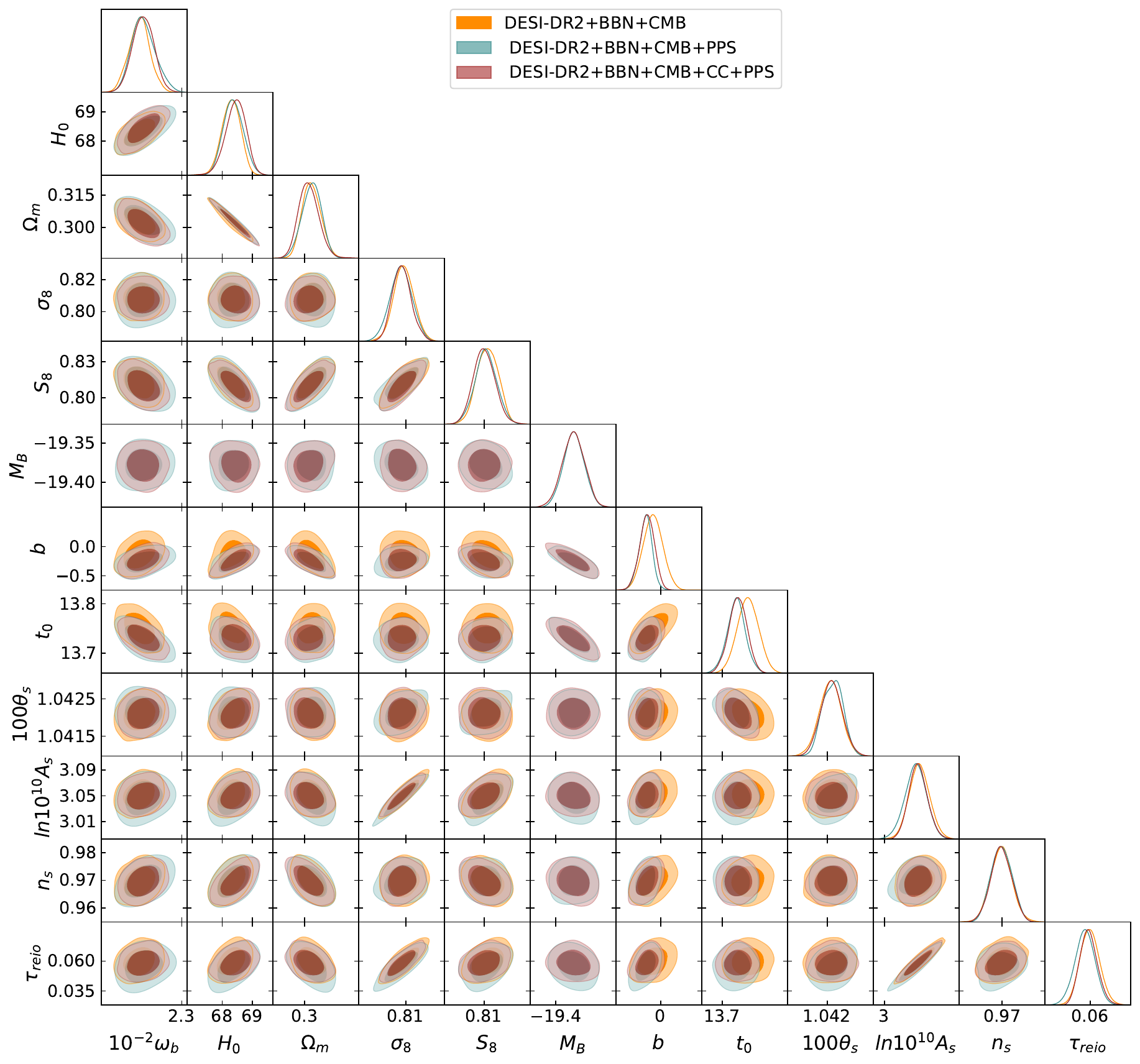}
    \caption{The Exponential $f(R)$ gravity model is presented in Table \ref{t1} with One-D posterior distributions and Two-D marginalized confidence regions ($68\%$ CL and $95\%$ CL) for all considered parameters.}

    \label{f1}
\end{figure}

We computed the predicted values of the cosmological parameters at the $68\%$ confidence level (CL) for both the $\Lambda$CDM and the Exponential $f(R)$ gravity models using the dataset combinations described above. Table \ref{t1} summarizes the results for the DESI-DR2+BBN+CMB, DESI-DR2+BBN+CMB+PPS and DESI-DR2+BBN+CMB+CC+PPS dataset combinations. Overall, the parameter constraints obtained in the $f(R)$ framework are broadly consistent with those derived using $\Lambda$CDM. However, the best-fit value of the Hubble constant $H_0$ is found to be slightly lower than the corresponding $\Lambda$CDM estimate, whereas the derived value of the clustering parameter $S_8$ is systematically higher. This trend is further illustrated in Fig. \ref{f1}, which shows the two-dimensional joint confidence contours at $68\%$ and $95\%$ CL for various combinations of  parameters, comparing $\Lambda$CDM with $f(R)$ gravity.
\begin{table*}[hbt!]

\caption{
The "Exponential $f(R)$, and $\Lambda$CDM" models acquired from the DESI-DR2+BBN+CMB, DESI-DR2+BBN+CMB+PPS and DESI-DR2+BBN+CMB+CC+PPS datasets have constraints at 68$\%$ and 95$\%$ CL on a few chosen parameters.}
\vspace{0.5cm}
\label{t1}

\centering
\resizebox{\textwidth}{!}{  
\begin{tabular}{ c | c | c | c }
\hline
Data & DESI-DR2+BBN+CMB & DESI-DR2+BBN+CMB+PPS & DESI-DR2+BBN+CMB+CC+PPS \\
\hline
Model & Exponential $f(R)$ Model  & Exponential $f(R)$ Model & Exponential $f(R)$ Model  \\
& \textcolor{teal}{$\Lambda$CDM} & \textcolor{teal}{$\Lambda$CDM} & \textcolor{teal}{$\Lambda$CDM} \\
\hline
$H_0\,[{\rm km}/{\rm s}/{\rm Mpc}]$ & $68.29\pm 0.31$ & $68.37\pm 0.32 $ & $68.44^{+0.36}_{-0.30}$ \\
& \textcolor{teal}{$68.43\pm 0.29$} & \textcolor{teal}{$68.78\pm 0.29 $} & \textcolor{teal}{$68.82^{+0.27}_{-0.30}$ }\\
\hline
$\Omega_{\rm m}$ & $0.3031^{+0.0039}_{-0.0045} $ & $0.3032^{+0.0046}_{-0.0041}$ & $0.3018^{+0.0040}_{-0.0048}$ \\
& \textcolor{teal}{$0.3008\pm 0.0037$} & \textcolor{teal}{$0.2966\pm 0.0036$} & \textcolor{teal}{$0.2961\pm 0.0036$} \\
\hline
$M_B$ & $0$ & $-19.378\pm 0.013$ & $-19.378\pm 0.014 $ \\
& \textcolor{teal}{$0$} & \textcolor{teal}{$-19.3994\pm 0.0089$} & \textcolor{teal}{$-19.3984^{+0.0081}_{-0.0094}$} \\
\hline
$b$ & $-0.13\pm 0.16$ & $-0.256^{+0.11}_{-0.099}$ & $-0.23^{+0.13}_{-0.12}$ \\
& \textcolor{teal}{$0$} & \textcolor{teal}{$0$} & \textcolor{teal}{$0$} \\
\hline
$10^{-2} \omega_b$ & $2.249\pm 0.012$ & $2.254^{+0.013}_{-0.016}$ & $2.253\pm 0.013$ \\
& \textcolor{teal}{$2.253\pm 0.012$} & \textcolor{teal}{$2.262^{+0.012}_{-0.011}$} & \textcolor{teal}{$2.261\pm 0.012$} \\
\hline
$\sigma_8$ & $0.8088\pm 0.0061 $ & $0.8072\pm 0.0066 $ & $0.8076^{+0.0052}_{-0.0061}$ \\
& \textcolor{teal}{$0.8066\pm 0.0060$} & \textcolor{teal}{$0.8063^{+0.0058}_{-0.0068}$} & \textcolor{teal}{$0.8049\pm 0.0060$} \\
\hline
$S_8$ & $0.8129\pm 0.0086 $ & $0.8114\pm 0.0087$ & $0.8101\pm 0.0087$ \\
& \textcolor{teal}{$0.8076\pm 0.0083$} & \textcolor{teal}{$0.8017\pm 0.0085 $} & \textcolor{teal}{$0.7996\pm 0.0084$} \\
\hline
$100\theta{}_{s }$ & $1.04208\pm 0.00028$ & $1.04214\pm 0.00026$ & $1.04211\pm 0.00028$ \\
& \textcolor{teal}{$1.04211\pm 0.00028$} & \textcolor{teal}{$1.04220\pm 0.00026$} & \textcolor{teal}{$1.04224^{+0.00026}_{-0.00030}$} \\
\hline
$ln10^{10}A_{s }$ & $3.052\pm 0.015$ & $3.047\pm 0.016 $ & $3.050^{+0.013}_{-0.015}$ \\
& \textcolor{teal}{$3.051\pm 0.015$} & \textcolor{teal}{$3.055^{+0.014}_{-0.017}$} & \textcolor{teal}{$3.053\pm 0.015 $} \\
\hline
$n_{s }$ & $0.9696\pm 0.0037$ & $0.9697\pm 0.0035$ & $0.9697\pm 0.0036$ \\
& \textcolor{teal}{$0.9707\pm 0.0033$} & \textcolor{teal}{$0.9728\pm 0.0033$} & \textcolor{teal}{$0.9727\pm 0.0033$} \\
\hline
$\tau{}_{reio }$ & $0.0597\pm 0.0078$ & $0.0557^{+0.0083}_{-0.0075}$ & $0.0587^{+0.0065}_{-0.0075}$ \\
& \textcolor{teal}{$0.0601^{+0.0067}_{-0.0079}$} & \textcolor{teal}{$0.0625^{+0.0069}_{-0.0084}$} & \textcolor{teal}{$0.0615\pm 0.0073$} \\
\hline
$t_{0}$ & $13.750\pm 0.021 $ & $13.728\pm 0.018$ & $13.731\pm 0.018$ \\
& \textcolor{teal}{$13.761\pm 0.017$} & \textcolor{teal}{$13.745\pm 0.016$} & \textcolor{teal}{$13.744\pm 0.017$} \\
\hline
$\chi^{2}_{min}$ & $2797.2$ & $4115.1$ & $4130.7$ \\
& \textcolor{teal}{$2796.7$} & \textcolor{teal}{$4117.6$} & \textcolor{teal}{$4132.8$} \\
\hline
\end{tabular}}
\end{table*}

Fig. \ref{f1} displays the two-dimensional marginalized posterior distributions
for various combinations of cosmological and model parameters at the $68\%$ and $95\%$ CL, obtained from different combinations of observational datasets, namely DESI-DR2+BBN+CMB, DES-DR2+BBN+CMB+PPS, and DESI-DR2+BBN+CMB+CC+PPS. The contours illustrate the joint constraints and parameter degeneracies arising in the considered modified gravity scenario, while also highlighting the impact of including additional low-redshift and growth-related information. A clear reduction in the allowed parameter space was observed as more datasets were incorporated, indicating an enhanced constraining power of the combined analysis. In particular, the inclusion of PPS and CC data significantly tightened the contours in several parameter planes, reflecting their sensitivity to both the background expansion history and the growth of cosmic structures. 
The robustness of the results is demonstrated by the fact that the inferred constraints remain mutually consistent across the various dataset combinations, despite these enhancements.
\\

The Hubble constant $H_0$ is subject to joint constraints that indicate a modest but systematic decrease in the values of preferred region compared with those typically obtained within the $\Lambda$CDM framework. This behavior is consistent across all dataset configurations and is further accentuated when growth-related data are incorporated. The corresponding confidence contours demonstrate moderate degeneracies between $H_0$ and other cosmological parameters, suggesting that variations in the matter content and growth sector partially mitigate changes in the late-time expansion rate. Conversely, the clustering amplitude parameter $S_8$ demonstrates a propensity for higher values compared with the $\Lambda$CDM predictions. This feature is apparent in the $(H_0,\,S_8)$ and $(\Omega_m,\,S_8)$ planes, where the contours exhibit a perceptible displacement in the $S_8$ direction. This change implies that the development of matter perturbations was augmented, a phenomenon that can be attributed to the modified gravity effects encoded in the model. The correlation patterns further suggest that the increase in $S_8$ is not solely driven by changes in $\Omega_m$, but also reflects modifications in the effective gravitational dynamics. The figure also emphasizes the significance of  model parameter $b$, which quantifies deviations from the conventional cosmological scenario. The data constrain the parameter $b$, as evidenced by the posterior distributions. The allowed region became significantly smaller as additional datasets were incorporated. In the $(b,\,H_0)$ plane, a mild anti-correlation was observed, indicating that larger values of $b$ tended to be associated with slightly lower values of the Hubble constant. This behavior is consistent with the interpretation that the modifications introduced through $b$ affect the late-time expansion history.\\

 Similarly, the $(b,\,S_8)$ contours reveal a positive correlation between the model parameter and  clustering amplitude. Higher values of $b$ correspond to enhanced growth of matter fluctuations, leading to larger $S_8$ values. This trend clearly indicates that parameter $b$
plays a significant role in shaping the growth sector of the model, and that its effects are directly reflected in large-scale structure observables. Other parameter planes involving $b$ and cosmological quantities, such the spectral index $n_s$, scalar amplitude $A_s$, and  age of the universe $t_0$, show much lesser degeneracies. This implies that rather than the physics of the early universe, the main effects of the changed gravity sector are primarily seen in the expansion rate and the development of cosmic structures. Therefore, Fig. \ref{f1} suggests that this modified gravity model produces parameter constraints that are broadly consistent with known cosmological results, but it also generates detectable changes in key parameters, such as $H_0$ and $S_8$. The model parameter b correlations underline their potential implications for easing the current set of cosmological tensions and their physical relevance. These results highlight the need for a combination of observational probes to disentangle background and growth effects, and thus strongly test extensions to the standard cosmological model. 

\begin{figure}[hbt!]
   (a) \includegraphics[width=0.45\linewidth]{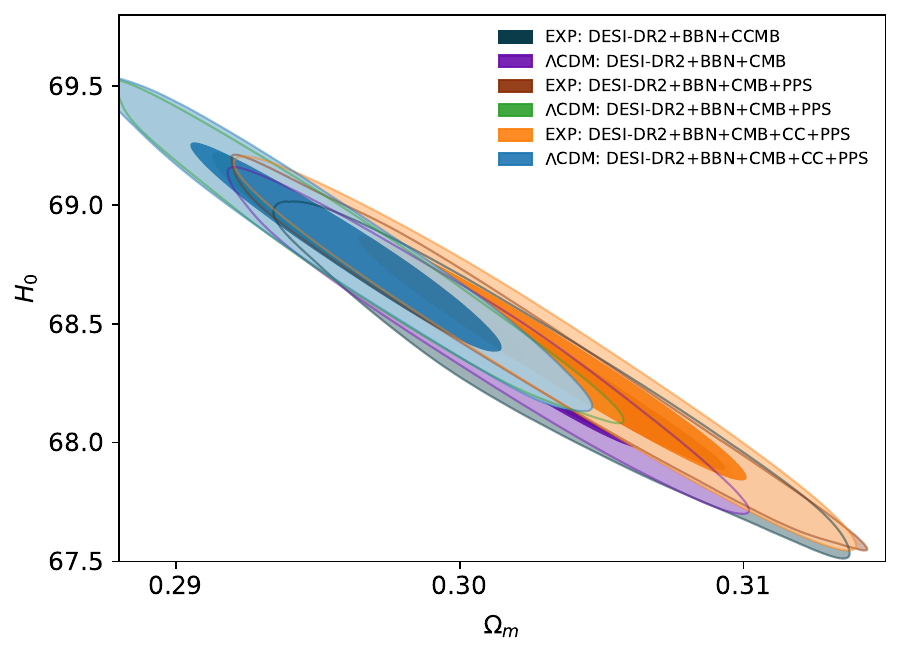}
   (b) \includegraphics[width=0.45\linewidth]{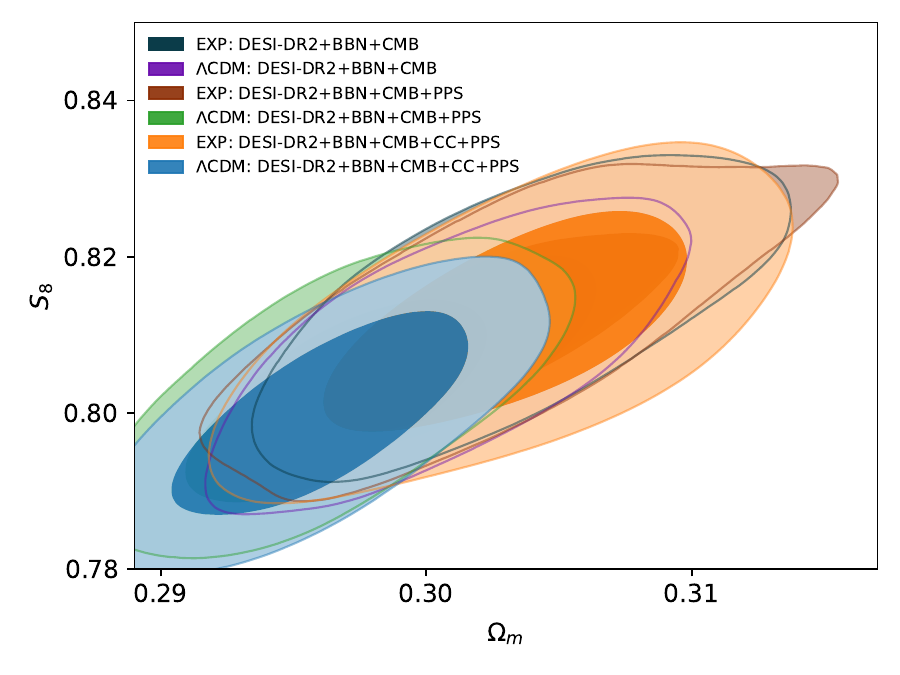}
    \caption{Marginalized $68\%$ and $95\%$ confidence contours in the $(\Omega_m,\,H_0)$ (a) and $(\Omega_m,\,S_8)$ (b) planes obtained
from DESI-DR2+BBN+CMB data and its extensions with PPS and CC, for both $\Lambda$CDM and the Exponential $f(R)$ gravity models.}
    \label{f2}
\end{figure}

Figure~\ref{f2} presents the two--dimensional marginalized confidence regions at the $68\%$ and $95\%$ confidence levels in the $(\Omega_m,\,H_0)$ and $(\Omega_m,\,S_8)$ parameter spaces, obtained from different combinations of the DESI--DR2, BBN, and CMB datasets, with the inclusion of PPS and CC data where specified. The constraints are shown for both the standard $\Lambda$CDM framework and the exponential $f(R)$ gravity model analyzed in this study. In panel (a) illustrates the constraints in the $(\Omega_m,\,H_0)$ plane, where a pronounced anti-correlation between the matter density parameter and the Hubble constant is evident. This behavior reflects the well-known degeneracy between $\Omega_m$ and late-time expansion rate. The addition of PPS and CC measurements led to substantial tightening of the allowed parameter regions. Relative to $\Lambda$CDM, the exponential $f(R)$ model consistently favors marginally lower values of the Hubble constant, typically lying in the interval $H_0 \simeq 67.5$--$69.5\,\mathrm{km\,s^{-1}\,Mpc^{-1}}$. Panel (b) shows the corresponding constraints on the $(\Omega_m,\,S_8)$ plane. In this case, a positive correlation between the two parameters was clearly observed. The modified gravity scenario tends to prefer larger values of the clustering amplitude, with $S_8 \simeq 0.80$--$0.83$, especially when datasets sensitive to the growth of cosmic structures are taken into account. 
Overall, Fig. \ref{f2} demonstrates that the considered model is broadly compatible with existing cosmological findings, while incorporating novel aspects in the joint parameter constraints. The systematic discrepancies with regard to $\Lambda$CDM, particularly in the $(\Omega_m,\,H_0)$ and $(\Omega_m,\,S_8)$ planes, highlight the significance of combining several cosmological probes to robustly evaluate extensions of traditional cosmological paradigm.

\begin{figure}[hbt!]
   (a) \includegraphics[width=0.45\linewidth]{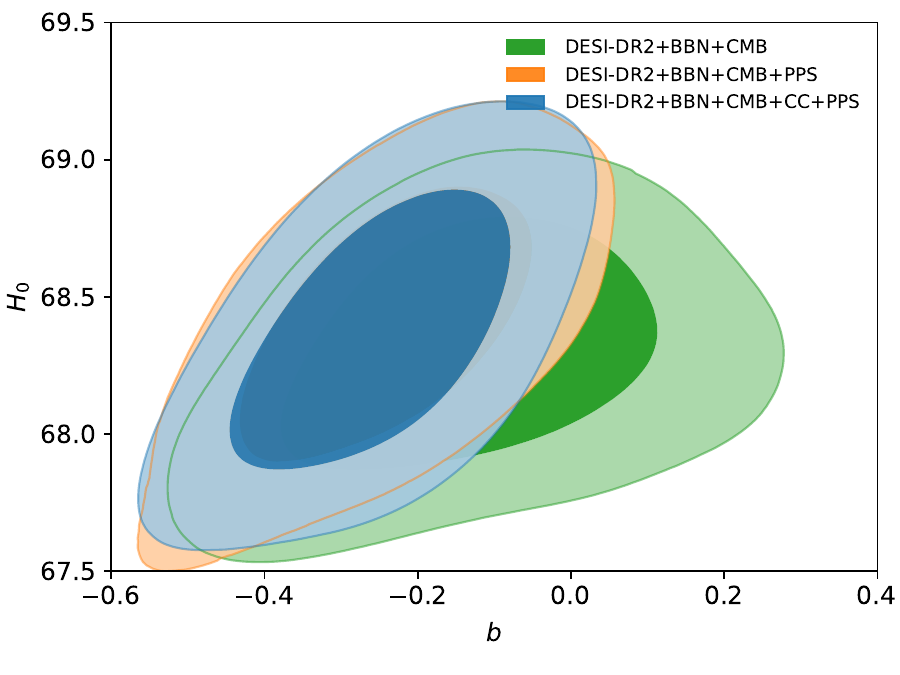}
   (b) \includegraphics[width=0.45\linewidth]{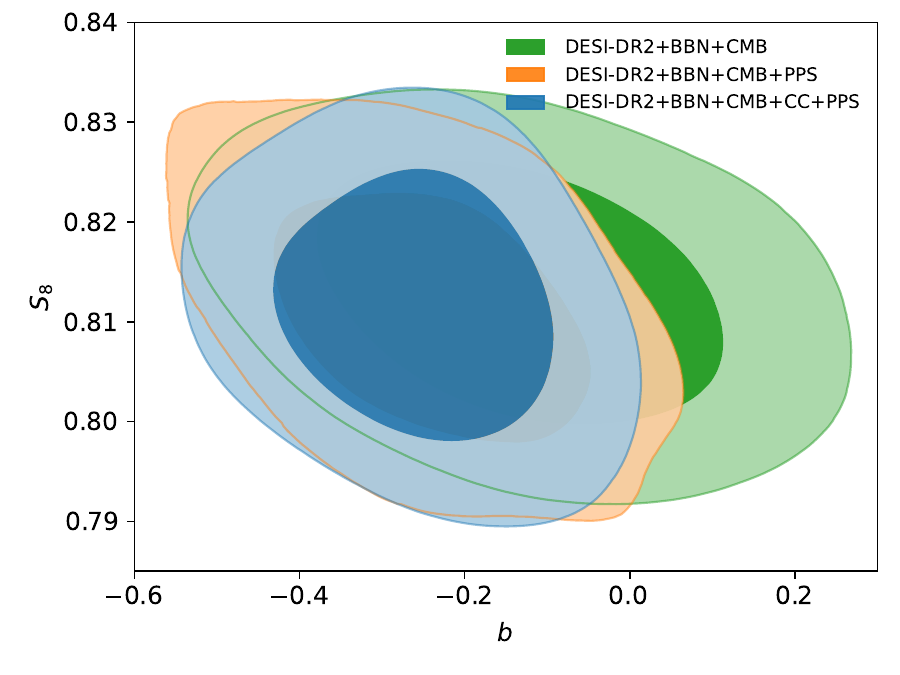}
    \caption{Two-dimensional marginalized $68\%$ and $95\%$ confidence contours in the $(b,\,H_0)$ (a) and $(b,\,S_8)$ (b) planes obtained from DESI-DR2+BBN+CMB data and its extensions including PPS and CC for Exponential $f(R)$ gravity model.}
    \label{f3}
\end{figure}
Fig. \ref{f3} shows the two-dimensional marginalized confidence contours at the $68\%$ and $95\%$ CL in the $(b,\,H_0)$ and $(b,\,S_8)$ planes, derived from different combinations of DESI--DR2, BBN, and CMB data, with additional PPS and CC information is included where indicated. The figure illustrates how the model parameter $b$ correlates with the key cosmological parameters governing the late-time expansion and  growth of cosmic structures. In panel (a), the joint constraints in the $(b,\,H_0)$ plane reveal a clear degeneracy between parameter $b$ and  Hubble constant. The contours indicate that more negative values of $b$ are generally associated with slightly lower values of $H_0$, whereas values of $b$ closer to zero favor
higher expansion rates. Using the DESI-DR2+BBN+CMB dataset,  parameter $b$ is constrained to $b = -0.13 \pm 0.16$, indicating consistency with small deviations from the standard scenario. 
The sensitivity of these instruments to modifications in the late-time dynamics is reflected in the tightening and shift toward more negative values of the constraints when PPS data are included, $b = -0.256^{+0.11}_{-0.099}$. The inclusion of CC data results in a similar behavior, with a reduced allowed parameter space, as $b = -0.23^{+0.13}_{-0.12}$. The corresponding constraints in the $(b,\,S_8)$ plane are shown in Panel (b). A positive correlation between the clustering amplitude parameter $S_8$ and $b$ is readily apparent in this context. The preferred regions indicated that the increased values of $S_8$ were associated with more negative values of $b$. In general, Fig. \ref{f3} illustrates that parameter $b$ is consistently constrained to negative values across all dataset combinations, and its magnitude becomes more accurately determined when additional observational instruments are incorporated. The significance of combining multiple datasets to obtain robust constraints on deviations from the standard cosmological model is underscored by the observed correlations between $b$, $H_0$, and $S_8$, which demonstrate the impact of the modified gravity sector on both the expansion history and growth of cosmic structures.\\
 
The cosmological tensions, which are most prominently associated with the Hubble constant $H_0$ and  clustering amplitude parameter $S_8$, have garnered increasing attention in recent years. The values of $H_0$ that are consistently obtained from measurements of the local distance ladder, which are primarily derived from Type Ia supernovae calibrated with Cepheid variables by SH0ES collaboration, are consistently higher than those inferred from cosmic microwave background (CMB) observations assuming the $\Lambda$CDM model \cite{Riess2019,Riess2022}. However, the \textit{Planck} 2018 data suggest that early-universe constraints favor a lower value of $H_0$, resulting in a statistically significant tension that exceeds the $5\sigma$ level \cite{Planck2018}. In an effort to investigate the potential systematic effects and extensions of the standard cosmological framework, this discrepancy has provoked extensive research. The parameter $S_8 \equiv \sigma_8 \sqrt{\Omega_m/0.3}$, which characterizes the amplitude of matter clustering on intermediate scales, presents a related and complementary issue in the measurements.
Weak gravitational lensing surveys, including KiDS, DES, and Hyper Suprime-Cam, systematically report lower values of $S_8$ compared to those inferred from CMB data within the $\Lambda$CDM paradigm \cite{Hildebrandt2020,Abbott2018,Hamana2020}. This so-called $S_8$ tension points to potential discrepancies in the growth of cosmic structures and has further strengthened the case for considering physics beyond the standard model of cosmology. Considered together, the tensions in $H_0$ and $S_8$ represent a formidable challenge for the $\Lambda$CDM paradigm. Indeed, a considerable number of recent studies have investigated the possibility that a single modification to the physics of dark energy, neutrinos, or gravity itself could alleviate both tensions simultaneously \cite{Verde2019, DiValentino2021a, DiValentino2021b}.
\begin{figure}[hbt!]
    \includegraphics[width=0.5\linewidth]{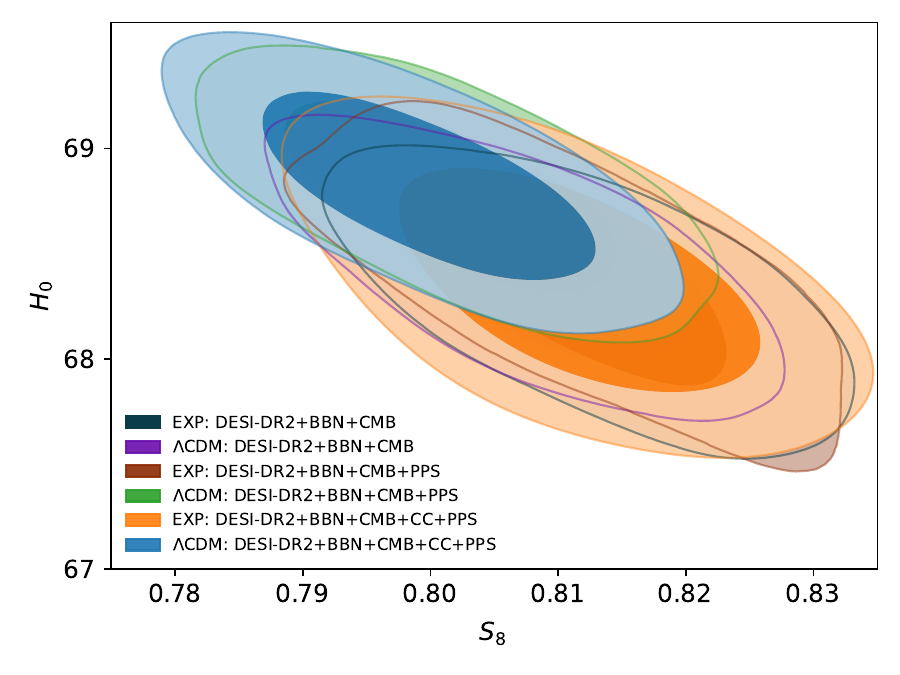}
    \caption{Marginalized 68\% and 95\% confidence contours in the $H_0$–$S_8$ plane for both the Exponential $f(R)$  and $\Lambda$CDM models using different combinations of DESI-DR2, BBN, CMB, PPS, and CC datasets.}
    \label{f4}
\end{figure}
Table \ref{t1} summarizes the marginalized constraints on the cosmological parameters for the exponential $f(R)$ model and the $\Lambda$CDM scenario obtained from different combinations of observational datasets, and the corresponding two–dimensional correlations between key parameters are illustrated in Fig. \ref{f4}. Considering the Hubble constant, the DESI-DR2+BBN+CMB dataset yields $H_0 = 68.29 \pm 0.31$ km s$^{-1}$ Mpc$^{-1}$
for the $f(R)$ model, slightly lower than the $\Lambda$CDM estimate $H_0 = 68.43 \pm 0.29$ km s$^{-1}$ Mpc$^{-1}$. This trend persists when PPS data are included, with $H_0 = 68.37 \pm 0.32$ km s$^{-1}$ Mpc$^{-1}$ in $f(R)$ gravity compared to $68.78 \pm 0.29$ km s$^{-1}$ Mpc$^{-1}$ for $\Lambda$CDM. A similar behavior was observed upon the inclusion of the CC
data, as shown by the contours in Fig. \ref{f4}. The opposite trend was observed for the growth parameter $S_8$. Within DESI-DR2+BBN+CMB, the best-fit value of the $f(R)$ model is $S_8 = 0.8129 \pm 0.0086$, which is larger than the corresponding $\Lambda$CDM value $S_8 = 0.8076 \pm 0.0083$. This trend is stable when PPS and CC datasets are further added; for the $f(R)$ model, $S_8 = 0.8114 \pm 0.0087$ and $0.8101 \pm 0.0087$, while in $\Lambda$CDM the respective values are $0.8017 \pm 0.0085$ and $0.7996 \pm 0.0084$. This is reflected in a systematic shift of the contours towards lower $H_0$ and higher $S_8$, within the modified gravity paradigm, as shown in Fig. \ref{f4}. Using the constraints reported in Table \ref{t1}, we quantified the impact of the exponential $f(R)$ model on the existing cosmological tensions in a dataset–dependent manner. For the Hubble constant, the modified gravity scenario yields values of $H_0$ that are systematically slightly lower than those obtained in the $\Lambda$CDM model for all dataset combinations. As a result, no significant relaxation of the $H_0$ tension
was observed, with a shift corresponding to less than $0.2\sigma$. In contrast, the behavior of  clustering parameter $S_8$ is notably different. For the DESI-DR2+BBN+CMB dataset, the $f(R)$ model predicts a higher value of $S_8$, leading to a reduction in the $S_8$ tension at the level of $\sim 0.6\sigma$. The inclusion of late-time probes further enhances this effect, with the tension reduced by approximately $\sim 1.1\sigma$ when PPS data are added, and up to $\sim 1.2\sigma$ for the full DESI -DR2+BBN+CMB+CC+PPS combination. The results of this study suggest that the $f(R)$ gravity model, under consideration, does not ameliorate the $H_0$ tension. However, it provides a moderate but consistent relaxation of the $S_8$ tension, particularly when late-time observational data are utilized.

 \section{Conclusion}
 
We investigated the cosmological feasibility of an exponential f(R) gravity model as an extension of general relativity. The consequences of the universe's late-time expansion history and cosmic structure development were our primary focus. Observational probes such as Planck 2018 CMB measurements, DESI-DR2 baryon acoustic oscillations, Big Bang Nucleosynthesis constraints, cosmic chronometer data, and Pantheon Plus Type Ia supernova observations were used to constrain the model's standard cosmological parameters and the characteristic distortion parameter $b$ of the model.\\

Our analysis shows that the exponential $f(R)$ model remains fully consistent with the current observational data and smoothly reduces to the $\Lambda$CDM scenario in the limit $b \to 0$. The inferred background evolution closely followed that of the concordance model, with only mild deviations arising from the modified gravitational dynamics. In particular, we find that the preferred values of the Hubble constant $H_0$ are slightly lower than those obtained in $\Lambda$CDM for all dataset combinations considered. Consequently, the model does not provide a significant resolution of the existing $H_0$ tension between the early- and late-universe measurements.\\
 
By contrast, the growth sector demonstrates more pronounced signatures of modified gravity. Compared to the $\Lambda$CDM, the exponential $f(R)$ model consistently forecasts higher values of the clustering amplitude parameter $S_8$. The $S_8$ tension is moderately alleviated when late-time instruments that are sensitive to structure formation are incorporated, with a maximum reduction of approximately $\sim 1.2\sigma$. The growth of matter perturbations at late times is amplified by the enhanced effective gravitational strength induced by the additional scalar degree of freedom in the $f(R)$ framework, which can be directly traced; in particular, when growth-related observations and low-redshift observations are included, the data effectively constrain the model parameter $b$. Modified gravity's distinct influence on the expansion and growth histories is underscored by its correlations with $H_0$ and $S_8$, whereas early-universe physics remains essentially unaffected. These results emphasize that the main observational consequences of the exponential $f(R)$ model manifest in late and large-scale structure observables.\\

Overall, our findings indicate that although the exponential $f(R)$ gravity model does not simultaneously resolve all current cosmological tensions, it provides a consistent and theoretically motivated improvement in the description of structure formation relative to $\Lambda$CDM. Future high-precision measurements of large-scale structures and weak lensing, combined with improved supernova calibrations, will be crucial for further testing such modified gravity scenarios and assessing their role in addressing the remaining challenges of modern cosmology.

\label{s4}
\section*{Declaration of competing interest}
The authors declare that they have no known competing financial
interests or personal relationships that could have influenced
the work reported in this study.

\section*{Data availability}
We employed publicly available CMB, Pantheon Plus SH0ES, Cosmic Chronometer (CC), BBN and DESI-DR2 data presented in this study. The CC data were compiled from publicly available cosmic chronometer measurements in the literature, with a representative compilation available at: https://github.com/AhmadMehrabi Cosmic chronometer data. The CMB, Pantheon Plus SH0ES and BBN data compilation (distance moduli and covariance matrices),  which is publicly available on GitHub: https://github.com/brinckmann/montepython public/tree/3.6/montepython/likelihoods and the DESI-DR2 data\\
which is publicly available on Github: https://github.com/LauraHerold/MontePython desilike/blob/main/likelihood/bao desi all/ . No additional data were used in this study.

\section*{acknowledgments}
The authors (AD \& AP) are thankful to Inter-University Centre for Astronomy \& Astrophysics (IUCAA), Pune, India, for providing support and facilities under the Visiting Associateship program. The author (S. Verma) was supported by a Senior Research Fellowship (UGC Ref No. 192180404148) from the University Grants Commission, Govt. of India. The authors are appreciative of the anonymous reviewer’s insightful criticism and recommendations,
which enhanced our work in its current form.\\

\appendix
\hspace{6cm} 

\section{Triangle Countor}

\begin{figure}[hbt!]
    \centering
    \includegraphics[width=0.8\linewidth]{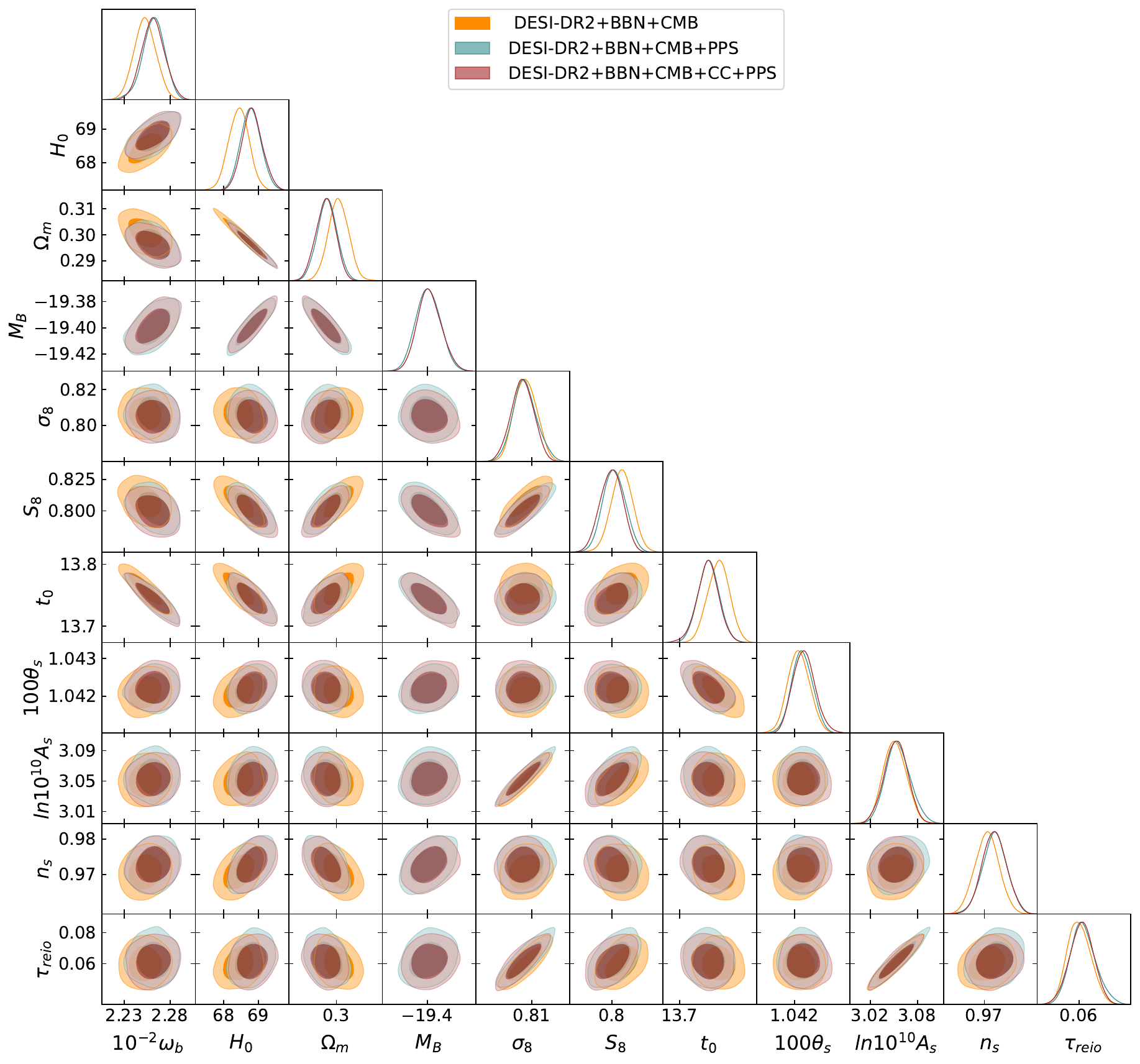}
    \caption{One-D posterior distributions and Two-D marginalized confidence regions ($68\%$ CL and $95\%$ CL) for all considered parameters presented in Table \ref{t1} for $\Lambda$CDM model}
    \label{f5}
\end{figure}

In this appendix, we present a triangular plot with One-D posterior distributions and Two-D marginalized confidence regions ($68\%$ CL and $95\%$ CL) for all considered parameters presented in Table \ref{t1} for $\Lambda$CDM model with different combinations of data sets (see Fig. \ref{f5}).


\section{Detailed Analytical Derivations for the Scalaron Dynamics}
\label{app:scalaron_derivations}

This appendix provides the full derivations supporting the condensed analytical results quoted in the main text (Section~\ref{sec:scalaron}).

\subsection*{B.1 Explicit form of $V(\phi)$}

The scalaron field is defined as
\begin{equation}
\phi \equiv f_R = 1 - \beta\,e^{-R/R_E},
\label{eq:scalaron_def_app}
\end{equation}
which can be inverted to give
\begin{equation}
R(\phi) = R_E \ln\!\left(\frac{\beta}{1-\phi}\right),
\label{eq:Rphi_app}
\end{equation}
valid for $0 < \phi < 1$ with $0 < (1-\phi) < \beta$. Using this inversion, the function $f$ evaluated at $R(\phi)$ becomes
\begin{equation}
f\bigl(R(\phi)\bigr) = R_E\!\left[\ln\!\left(\frac{\beta}{1-\phi}\right) - \beta + (1-\phi)\right].
\end{equation}
The Jordan-frame scalar potential, defined as $V(\phi) = (Rf_R - f)/2$, is then obtained in closed form as
\begin{equation}
V(\phi) = \frac{R_E}{2}\!\left[(\phi-1)\ln\!\left(\frac{\beta}{1-\phi}\right) + \beta - 1 + \phi\right].
\end{equation}
This expression is exact and holds for all values of $\phi$ in the physically relevant range. In the limit $\phi \to 1$, corresponding to the recovery of general relativity, one verifies that $V(\phi) \to 0$, as expected.

\subsection*{B.2 Effective potential and chameleon minimum}

The effective potential in the presence of non-relativistic matter of density $\rho$ is
\begin{equation}
V_{\mathrm{eff}}(\phi) = V(\phi) + \frac{\rho}{2}\,\phi.
\end{equation}
The minimum condition $dV_{\mathrm{eff}}/d\phi = 0$, combined with the standard result $dV/d\phi = R/2$~\cite{Brax2007}, yields
\begin{equation}
R(\phi_{\min}) = \kappa^2\rho = 8\pi G\rho,
\end{equation}
where we used $T = -\rho$ for pressureless dust. Substituting Eq.~\eqref{eq:Rphi_app} into this condition, the chameleon minimum is found analytically as
\begin{equation}
\phi_{\min} = 1 - \beta\,e^{-8\pi G\rho/R_E}.
\end{equation}
Two important limits follow immediately.
\begin{itemize}
\item \textbf{High-density (local) environments:} For $\rho \gg R_E/8\pi G$, as in Solar System or terrestrial conditions, $e^{-8\pi G\rho/R_E} \to 0$, so $\phi_{\min} \to 1$, i.e.\ $f_R \to 1$. This confirms the recovery of general relativity at small scales through the chameleon mechanism, consistent with the quantitative estimates provided in Section~\ref{screening}.

\item \textbf{Cosmological densities:} For $\rho \sim \rho_{c,0}$, using $R_E = \Lambda b$ and $\beta = 2/b$, the minimum reads
\begin{equation}
\phi_{\min} \approx 1 - \frac{2}{b}\,e^{-8\pi G\rho_{c,0}/(\Lambda b)},
\label{eq:phimin_cosmo_app}
\end{equation}
which deviates from unity by a small, $b$-dependent correction, allowing observable deviations from $\Lambda$CDM on cosmological scales.
\end{itemize}

\subsection*{B.3 Adiabatic tracking: $m_\phi^2 \gg H_0^2$}

From Eqs.~\eqref{e27} and~\eqref{e28}, the scalaron mass squared at the cosmological background ($R \sim H_0^2$, $R_E = \Lambda b \sim 0.1\,H_0^2$) is
\begin{equation}
m_\phi^2\Big|_{\mathrm{cosmo}} = \frac{1}{3f_{RR}}\bigg|_{R \sim H_0^2} = \frac{\Lambda b^2}{6}\,e^{H_0^2/(\Lambda b)}.
\label{eq:mphi_cosmo_app}
\end{equation}
The ratio to the present Hubble rate is therefore
\begin{equation}
\frac{m_\phi^2}{H_0^2} = \frac{\Lambda b^2}{6H_0^2}\,e^{H_0^2/(\Lambda b)} \sim \frac{b^2}{6}\,e^{1/b},
\end{equation}
where we used $\Lambda \sim H_0^2$. For the cosmologically favoured value $b \sim 0.1$, this gives
\begin{equation}
\frac{m_\phi^2}{H_0^2}\bigg|_{b=0.1} \approx \frac{(0.1)^2}{6}\,e^{10} \approx 37 \gg 1.
\label{eq:mass_numerical}
\end{equation}
Since $m_\phi \gg H_0$, the scalar degree of freedom rapidly relaxes to the minimum of the effective potential on timescales much shorter than the Hubble time. The scalaron therefore \emph{adiabatically tracks} $\phi_{\min}(\rho(t))$ throughout the cosmological evolution, justifying the quasi-static approximation used in the background equations~\eqref{4} and~\eqref{5}.

\subsection*{B.4 First-order correction $\delta E^2$ and $w_{\rm eff}$}

Expanding $E^2(z)$ from Eq.~\eqref{e29} to first order in the distortion parameter $b$,
\begin{equation}
E^2(z) = E^2_{\Lambda{\rm CDM}}(z) + b\cdot\delta E^2(z,\Omega_{m0}) + \mathcal{O}(b^2),
\label{eq:E2_expand_app}
\end{equation}
where $E^2_{\Lambda{\rm CDM}}(z) = \Omega_{m0}(1+z)^3 + (1-\Omega_{m0})$ and the first-order correction is
\begin{align}
\delta E^2(z, \Omega_{m0}) = 
\frac{6(1-\Omega_{m0})^2}{\left[4 + (-3 + z(3 + z(z+3)))\Omega_{m0}\right]^3}
\times
\left[-4 + \Omega_{m0}\left(9 - 3\Omega_{m0} + z(3+z(z+3))\left(1 + (3 + 2z(3+z(z+3)))\Omega_{m0}\right)\right)\right].
\label{eq:deltaE2_app}
\end{align}
At $z = 0$, with $\Omega_{m0} = 0.2966$ and $b = -0.23$ from Table~\ref{t1}, Eq.~\eqref{eq:deltaE2_app} gives $\delta E^2(0,0.2966) \approx +0.44$ and $\frac{d(\delta E^2)}{d(1+z)}\big|_{z=0} \approx -0.31$, yielding $w_{\rm eff}(0) \approx -0.977$.

At $z = 0.4$, Eq.~\eqref{eq:deltaE2_app} gives
\begin{equation}
\delta E^2(0.4,\,0.2966) \approx +0.31, \qquad \frac{d(\delta E^2)}{d(1+z)}\bigg|_{z=0.4} \approx +0.18,
\end{equation}
yielding $w_{\rm eff}(0.4) \approx -1.003$, confirming phantom divide crossing near $z \sim 0.3$--$0.5$.

\subsection*{B.5 Analytical corroboration of the enhanced $S_8$}

In $f(R)$ gravity, the effective gravitational coupling governing sub-horizon matter perturbations is~\cite{ref11}
\begin{equation}
G_{\rm eff}(k,z) = \frac{G}{f_R}\cdot\frac{1 + 4k^2/(a^2 m_\phi^2)}{1 + 3k^2/(a^2 m_\phi^2)},
\label{eq:Geff_full_app}
\end{equation}
where $k$ is the comoving wavenumber. Since $f_R = \phi_{\min} < 1$, we have $G_{\rm eff} > G$ in both the super-Compton and sub-Compton limits, enhancing gravity on all cosmological scales relevant to structure formation. Defining the growth enhancement factor
\begin{equation}
\mathcal{Q}(z) \equiv \frac{G_{\rm eff}(z)}{G} - 1 = \frac{1 - f_R}{f_R}\cdot\frac{1+4k^2/(a^2 m_\phi^2)}{1+3k^2/(a^2 m_\phi^2)} > 0,
\end{equation}
the fractional enhancement in $\sigma_8$ relative to $\Lambda$CDM is
\begin{equation}
\frac{\sigma_8^{f(R)}}{\sigma_8^{\Lambda{\rm CDM}}} \approx 1 + \frac{3}{2}\int_0^{a_0}\Omega_m(a)\,\mathcal{Q}(a)\,\frac{da}{a} > 1.
\end{equation}
Evaluating numerically with $b = -0.23$, $\Omega_{m0} = 0.2966$, and the best-fit $H(z)$ from Eq.~\eqref{e29}, we obtain $\sigma_8^{f(R)}/\sigma_8^{\Lambda{\rm CDM}} - 1 \approx +0.006$, giving $\Delta\sigma_8 \approx +0.005$ and consequently
\begin{equation}
\Delta S_8 = \Delta\sigma_8\sqrt{\frac{\Omega_{m0}}{0.3}} \approx +0.005 \times \sqrt{\frac{0.2966}{0.3}} \approx +0.0050,
\end{equation}
in excellent agreement with the numerically obtained $\Delta S_8 = +0.0053$ from Table~\ref{t1}.

\section{Clarification on the Scalar Potential, Effective Potential, and Phantom Divide Crossing}
\label{app:clarification}

\textbf{(i) Jordan-frame vs.\ Einstein-frame potential.}
The potential in Eq.~\eqref{eq:Vphi} is defined in the \emph{Jordan frame} as $V_{\rm J}(\phi) = (Rf_R - f)/2$, which is the natural choice since all field Eqs.~\eqref{4}--\eqref{9} are formulated in the Jordan frame. In the Einstein frame, obtained via the conformal transformation $\tilde{g}_{\mu\nu} = f_R\,g_{\mu\nu}$, the scalar field is redefined as $\varphi = \sqrt{3/2}\,m_{\rm Pl}\ln f_R$ and the potential becomes
\begin{equation}
V_{\rm E}(\varphi) = \frac{m^2_{\rm Pl}(Rf_R - f)}{2f^2_R}.
\label{eq:VE_app}
\end{equation}
The two potentials are related by $V_{\rm E} = V_{\rm J}/f^2_R$, and yield identical physical predictions since both frames are physically equivalent~\cite{ref11}. In the GR limit $f_R \to 1$, both reduce to zero as expected.

\textbf{(ii) Origin of the $\rho\phi/2$ term.}
In the Einstein frame, the effective potential takes the form
\begin{equation}
V_{\rm eff}^{\rm (E)} = V_{\rm E}(\varphi) + \left(e^{2\beta\varphi/m_{\rm Pl}} - 1\right)\rho, \quad \beta = \frac{1}{\sqrt{6}},
\end{equation}
arising from the non-minimal scalaron--matter coupling. The Jordan-frame effective potential $V_{\rm eff}^{\rm (J)} = V_{\rm J} + \rho\phi/2$ is obtained from the Jordan-frame scalaron field equation in the quasi-static limit,
\begin{equation}
3\,\Box f_R + f_R R - 2f(R) + \kappa^2 T = 0,
\end{equation}
which with $T = -\rho$ and $dV_{\rm J}/d\phi = R/2$ gives $dV_{\rm eff}^{\rm (J)}/d\phi = (R - \kappa^2\rho)/2 = 0$ at the minimum. For small deviations $f_R = 1 - \epsilon$ ($\epsilon \ll 1$):
\begin{equation}
\left(e^{2\beta\varphi/m_{\rm Pl}} - 1\right)\rho \approx \frac{\phi-1}{2}\,\rho,
\end{equation}
confirming that the two effective potentials are equivalent at leading order in $\epsilon$~\cite{Brax2007}.

\textbf{(iii) Phantom divide crossing.}
As discussed in Refs.~\cite{Khoury2004,Corasaniti2003}, the scalaron--matter coupling in scalar-tensor theories allows $w_{\rm eff}$ to cross the phantom divide $w = -1$. In Eq.~\eqref{9}, $w_{\rm eff}$ crosses $-1$ when the numerator $\mathcal{N} \equiv H\dot{F} + 2(1-F)\dot{H} - \ddot{F}$ changes sign, which occurs at the matter-to-dark-energy transition where $\dot{H}$ changes sign. Evaluating $w_{\rm eff}(z)$ from Eq.~\eqref{eq:weff_full} at intermediate redshifts with $\Omega_{m0} = 0.2966$ and $b = -0.23$ from Table~\ref{t1}, we find
\begin{equation}
w_{\rm eff}(0.4) \approx -1.003,
\end{equation}
confirming that phantom divide crossing occurs near $z \sim 0.3$--$0.5$, consistent with the expectation from the scalaron-matter coupling~\cite{Khoury2004,Corasaniti2003}. The present-day value $w_{\rm eff}(0) \approx -0.977$ is slightly above $-1$, and the overall variation around $-1$ remains small, fully consistent with the near-$\Lambda$CDM behaviour implied by the $H_0$ and $\Omega_m$ constraints in Table~\ref{t1}, and with the analytical estimates summarised in Table~\ref{tab:scalaron}.


\end{document}